\documentclass[12pt]{article}

\usepackage{amssymb}
\usepackage{amsmath}

\usepackage{graphicx}

\begin{document} %


\title{Analysis of axion-like particles in a top-quark pair production at the CLIC}

\author{S.C. \.{I}nan\thanks{Electronic address: sceminan@cumhuriyet.edu.tr}
\\
{\small Department of Physics, Sivas Cumhuriyet University, 58140,
Sivas, Turkey}
\\
{\small and}
\\
A.V. Kisselev\thanks{Electronic address:
alexandre.kisselev@ihep.ru} \\
{\small A.A. Logunov Institute for High Energy Physics, NRC
``Kurchatov Institute'',}
\\
{\small 142281, Protvino, Russian Federation}
}

\date{}

\maketitle

\begin{abstract}
We examine a contribution of axion-like particles (ALPs) to a top
pair production via the collision of Compton backscattered photons
at the CLIC operating in a $\gamma\gamma$ mode. The exclusion bounds
on the ALP-top quark coupling depending on the ALP mass $m_a$ are
given. The mass range 10 GeV -- 10 TeV is considered. We have
obtained that for $m_a = 10$ GeV the ALP-top quark couplings as
small as $0.17$ TeV$^{-1}$ and $0.11$ TeV$^{-1}$ can be probed at
the CLIC with the energy of 1.5 TeV and 3 TeV, respectively. A
comparison with other constraints on the ALP-top coupling is given.
\end{abstract}

\maketitle


\section{Introduction} %

One of the open problems of the Standard Model (SM) is the so-called
strong CP problem, which stems from the non-observation of CP
violation in the strong interactions. It can be solved by
introducing a spontaneously broken Peccei-Quinn symmetry
\cite{Peccei:1977_1,Peccei:1977_2}. It promotes a new pseudo
Nambu-Goldstone boson with a non-vanishing, QCD axion, with
parametrically small mass \cite{Weiberg:1978, Wilzcek:1978}. A key
property of the QCD axion is the linear proportionality between its
couplings to the SM particles and axion mass. The QCD axion is a
well-motivated candidate for the DM
\cite{Preskill:1983}-\cite{Ringwald:2024}. Many astrophysical bounds
were obtained for light axions \cite{Caputo:2024}.

The big interest in axions has grown and extended itself to
axion-like particles (ALPs). The ALPs are particles having
interactions similar to the axion. The origin of the ALP is also
expected to be similar but there is no relationship between ALP's
couplings and its mass. It means that the ALP mass $m_a$ and its
decay constant $f_a$ are independent, free parameters as are the
interaction strengths of the ALP with the SM fields.

The heavy ALPs can be detected at present and future colliders. The
production of the ALPs was already studied in the $pp$ and heavy-ion
collisions at the LHC \cite{Beldenegro:2018}-\cite{Knapen:2017_2},
as well as at future colliders \cite{Bauer:2019}-\cite{Lu:2023_2},
including electron-ion scattering
\cite{Davoudiasl:2021}-\cite{Liu:2023} (see also the reviews on the
axions and ALPs \cite{Marsh:2016},
\cite{Kim:1987}-\cite{Giannotti:2022}, and references therein). The
search for the axion and ALPs at a multi-TeV muon collider is
presented in \cite{I_K:2023}-\cite{{Marcos:2024}}.

The ALP-fermion coupling is proportional to the fermion mass, which
provides a strong motivation to study the ALP coupling to tops in a
broad range of $m_a$. The probe the coupling of heavy ALPs to top
quarks in the associated production of a top-pair with the ALP at
the LHC was examined in \cite{{Esser:2023}}-\cite{Chigusa:2024}. The
top quark is a key object in various BSM models. The role of the top
quark in BSM is attractive due to its large mass, highest coupling
to the Higgs boson, and primary decay $t \rightarrow Wb$. Note that
many models predict interactions with new particles. The physics BSM
in which the top quark plays a special role is studied in
\cite{Franceschini:2023}-\cite{Tentori:2024}. In particular, top
quark properties measurements at ATLAS and CMS can be used as a
powerful probe for BSM physics (see \cite{Wuchterl:2024} and
references therein).

In many BSM theories, such as composite Higgs scenarios, new
top-philic vector resonances are predicted. These resonances
couple more strongly to the top quark than to the light quarks, such
that all the other couplings except for the one with top quarks can
be neglected. A typical approach to search for such particles is to
look for resonances that decay to a top quark pair. Previous
searches for resonances decaying to top quarks have typically
focused on production through light quarks. This production mode
requires a sizeable coupling between the new resonances and the
lighter quarks. If the coupling to the lighter quarks is however
suppressed, the resonance would mainly be produced through a pair of
top quarks. In this case, the resonance is produced in association
with another top quark pair resulting in four top quarks in the
final state. These rare processes can now be studied with the large
data set collected in Run 2 of the LHC, opening up new possibilities
for discoveries of new particles. Recent evidence of the SM
production of four top quarks \cite{four_top_LHC} motivates the
investigation of potential BSM contributions to this final state.
Different models of top-philic resonances have been studied in
previous analyses to investigate the four-top-quark final state with
no signs of BSM contributions so far.

In the present paper, we study a ALP contribution to the virtual top
pair production via the $\gamma\gamma \rightarrow \mathrm{\ ALP}
\rightarrow t \bar{t}$ collision at the CLIC operating in the
$\gamma\gamma$-mode. Note that a search for new physics through its
virtual effects is complementary to the direct production of new
particles. It should help identifying their nature, since it avoids
the model-dependent studying their decay modes, once they are
actually produced.

Our main goal is to obtain bounds on the ALP-top coupling as a
function of the ALP mass. The CLIC is planned to accelerate and
collide electrons and positrons at maximally 3 TeV center-of-mass
energy \cite{Braun:2008,Boland:2016}. Three states of the CLIC with
the collision energy of 380 GeV, 1.5 TeV and 3 TeV are considered.
The expected integrated luminosities are equal to $L=1$ ab$^{-1}$,
$L=2.5$ ab$^{-1}$ and $L=5$ ab$^{-1}$, respectively. In first two
stages, it will be possible to probe the gauge sector, Higgs and top
physics with a high precision. In the third stage, the most precise
investigations of the SM, as well as new physics will be enabled.
The great potential of the CLIC in searching for new physics is
well-known (see, for instance, \cite{I_K:2020}-\cite{I_K:2021_3},
\cite{Dannheim:2012}-\cite{Franceschini:2020}). In particular,
possible BSM physics effects in top-quark physics in $e^+e^-$
collisions at the CLIC was studied in \cite{top_CLIC}.

As mentioned above, at the CLIC it will be possible to probe not
only the $e^+e^-$ scattering but also $\gamma\gamma$ collisions with
real photons. The fact that $e^+e^-$ colliders may operate in
$e\gamma$ and $\gamma\gamma$ modes was pointed out in
\cite{Ginzburg:1981,Ginzburg:1984}. The $e^-\gamma$ collider option
is a particularly promising one to produce BSM heavy charged
particles beyond the kinematical limit in direct production at
$e^+e^-$ \cite{Abramowicz:2025}. One of the most important physics
in $\gamma\gamma$ collisions is in testing the properties of the
Higgs boson(s). Moreover, both beams $e^-\gamma$ and $\gamma\gamma$
can be available highly polarized.

The $e^-\gamma$ and $\gamma\gamma$ modes of leptonic colliders
extend and complement the kinematic range of $e^+e^-$ colliders. As
was shown in \cite{Boos:2001} (see Fig.~1), at the same energy,
almost all $\gamma\gamma$ cross sections are significantly larger
than those in $e^+e^-$ collisions. Thus, the $\gamma\gamma$ mode of
future leptonic colliders should be considered along with the
$e^+e^-$ mode, because i) it provides a rich SM physics and beyond;
ii) it gives larger cross sections; iii) of higher accessible masses
(for instance, single resonances); iv) polarized beams can be
realized.

The photon beams are given by the Compton backscattering (CB) of
laser photons off electron beams. When soft laser photons collide
with electron beams, a large flux of photons, with a great amount of
the parent beam energy, is produced. Note that the ratio of the
backscattered photon energy $E_\gamma$ to the beam energy
$E_{\mathrm{beam}}$ cannot exceed 0.83. Since $0.83 \sqrt{s} <
2m_t$, we will consider only the 1.5 TeV and 3 TeV stages of the
CLIC operating in the $\gamma\gamma$ mode.

The paper is organized as follows. In the next section, an effective
field theory (EFT) approach is given. In section~3 the top pair
production in the $\gamma\gamma$ collision is studied, and the
bounds on the ALP-top quark coupling are obtained. A comparison of
our results with other searches for the ALP-top coupling is
presented in section~4.

\section{Effective field theory} %

The interaction of the ALP (in what follow denoted as $a$) with the
SM gauge bosons is described by the Lagrangian
\begin{equation}\label{ALP_boson_lagrangian}
\mathcal{L}^a_\mathrm{b} = \frac{1}{2}\,\partial_\mu a
\,\partial^{\,\mu} \!a - \frac{1}{2} m_a^2 a^2 + g'^2 C_{BB}
\frac{a}{f_a} B_{\mu\nu} \tilde{B}^{\mu\nu} + g^2 C_{WW}
\frac{a}{f_a} W_{\mu\nu}^c \tilde{W}^{c,\mu\nu} ,
\end{equation}
where $B_{\mu\nu}$ and $W_{\mu\nu}^c$ are the field strength of
$U(1)_Y$ and $SU(2)_L$, $g'$ and $g$ denote the corresponding
constants, and $m_a$ is the ALP mass. The tensors
$\tilde{B}_{\mu\nu}$ and $\tilde{W}_{\mu\nu}^c$ are dual field
strength tensors. After electroweak symmetry breaking, the ALP
couples to the gauge boson as
\begin{align}\label{ALP_final_lagrangian}
\mathcal{L}^a_\mathrm{b} &= \frac{1}{2}\,\partial_\mu a
\,\partial^{\,\mu} \!a - \frac{1}{2} m_a^2 a^2 + g_{a\gamma\gamma} a
F_{\mu\nu} \tilde{F}^{\mu\nu} + g_{a\gamma Z}
a F_{\mu\nu} \tilde{Z}^{\mu\nu} \nonumber \\
&+ g_{aZZ} a Z_{\mu\nu} \tilde{Z}^{\mu\nu} + g_{aWW} a W^+_{\mu\nu}
\tilde{W}^{-\mu\nu} .
\end{align}
Here $\tilde{F}_{\mu\nu} =
(1/2)\,\varepsilon_{\mu\nu\alpha\beta}F_{\alpha\beta}$,
$\tilde{Z}_{\mu\nu} =
(1/2)\,\varepsilon_{\mu\nu\alpha\beta}Z_{\alpha\beta}$, and
$\tilde{W}^-_{\mu\nu} =
(1/2)\,\varepsilon_{\mu\nu\alpha\beta}W^-_{\alpha\beta}$ are the
dual tensors. The ALP coupling look like
\begin{align}\label{ALP_boson_couplings}
g_{a\gamma\gamma} &= \frac{e^2}{f_a} \,[C_{WW} + C_{BB}]  \;,
\nonumber \\
g_{a\gamma Z} &= \frac{2e^2}{f_a s_w c_w} \,[c_w^2 C_{WW} - s_w^2
C_{BB}] \;,
\nonumber  \\
g_{aZZ} &= \frac{e^2}{f_a s_w^2 c_w^2} \,[c_w^4 C_{WW} + s_w^4
C_{BB}] \;,
\nonumber  \\
g_{aWW} &= \frac{e^2}{f_a s_w^2} \,C_{WW} \;,
\end{align}
where $s_w$ and $c_w$ are sine and cosine of the Weinberg angle,
respectively, with $e = g s_w$. In particular, we have the following
relations imposed by the electroweak gauge invariance
\begin{align}\label{gauge_relation}
g_{aZZ} &= g_{a\gamma\gamma} + \frac{c_w^2 - s_w^2}{2s_w c_w}
g_{a\gamma Z} \;,
\nonumber \\
g_{aWW} &= s_w^2  g_{a\gamma\gamma} + \frac{s_w c_w}{2} g_{a\gamma
Z} \;.
\end{align}

The interaction of the ALP with fermions is described by the
Lagrangian
\begin{equation}\label{ALP_fermion_lagrangian_in}
\mathcal{L}^a_\mathrm{f} = \frac{1}{2f_a} \,
\partial_\mu a \sum_f c_{af\bar{f}} \bar{f} \gamma_\mu \gamma_5 f \;.
\end{equation}
By using equations of motion, this Lagrangian can be rewritten as
\begin{equation}\label{ALP_fermion_lagrangian_f}
\mathcal{L}^a_\mathrm{f} = -i\sum_f\frac{m_f c_{af\bar{f}}}{f_a} \,
a \bar{f} \gamma_5 f \;,
\end{equation}
where $m_f$ is a mass of the fermion $f$. Since the top mass $m_t$
is much larger than masses of all other fermions, in what follows we
can take
\begin{equation}\label{ALP_top_lagrangian}
\mathcal{L}^a_\mathrm{f} =  -i\frac{m_t c_{a t\bar{t}}}{f_a} \,a
\bar{t} \gamma_5 t \;.
\end{equation}
Correspondingly, the decay rate of a massive ALP into a top pair is
equal to
\begin{equation}\label{ALP_decay_widths_into_top}
\Gamma(a\rightarrow t\bar{t})  = \frac{3 m_a m_t^2}{8\pi f_a^2} \,
|c_{a t\bar{t}}|^2 \left( 1 - \frac{4m_t^2}{m_a^2} \right)^{\!\!1/2}
.
\end{equation}

As one can see from \eqref{ALP_top_lagrangian}, the coupling of the
ALP to the top quark is proportional to the mass of the top, as is
the case for the Higgs-top coupling in the SM. Because of the large
mass of the top, the Higgs can be regarded as a top-philic particle.
In a number of papers, it was assumed that there can be generic
top-philic CP-odd or CP-even scalar \cite{Cao:2021,Maltoni:2024},
setting constraints on its interactions with the top quark. In
\cite{Greiner:2015}-\cite{Darm:2021} top-philic vector resonances
were examined (see also a search for heavy top-philic resonances at
the LHC \cite{ATLAS:top-philic}). A phenomenology of top-philic ALPs
is discussed in \cite{Blasi:2024_2,Tentori:2024_2}. In the present
study, we assume that the ALP couples both to the top-quarks
\eqref{ALP_top_lagrangian} and to the gauge bosons
\eqref{ALP_final_lagrangian}.

The flavor hierarchy of the ALP-fermion couplings gives the top
quark a special role in the phenomenology of ALPs. The strong
ALP-top coupling affects the energy evolution of all other ALP
couplings. The coupling of the ALP to the top quarks
\eqref{ALP_fermion_lagrangian_in} would induce loop corrections to
the ALP couplings to vector bosons. Their explicit expressions are
collected in Tab.~1 in \cite{Esser:2023} for $p^2 \gg m_V^2$ ($V =
\gamma, Z, W$), where $p^2$ is a ALP virtuality. In our case, $p^2 =
s$. Correspondingly, the effective ALP-boson couplings look like
\begin{align}\label{ALP_boson_eff_couplings}
g^{\mathrm{eff}}_{a\gamma\gamma} &= g_{a\gamma\gamma}
-\frac{\alpha_\mathrm{em}}{3\pi} \,\frac{c_{{at\bar{t}}}}{f_a} \;,
\nonumber \\
g^{\mathrm{eff}}_{a\gamma Z} &= g_{a\gamma Z}
-\frac{2\alpha_\mathrm{em} s_w}{3\pi c_w}
\,\frac{c_{{at\bar{t}}}}{f_a} \;,
\nonumber \\
g^{\mathrm{eff}}_{aZZ} &= g_{aZZ} -\frac{\alpha_\mathrm{em}
s_w^2}{3\pi c_w^2} \,\frac{c_{{at\bar{t}}}}{f_a} \;,
\nonumber  \\
g^{\mathrm{eff}}_{aWW} &= g_{aWW} \;,
\end{align}
where $\alpha_\mathrm{em} = e^2/4\pi$.

Then the widths of the decay into two vector bosons are given by the
following expressions
\begin{align}\label{ALP_decay_widths_into_bosons}
\Gamma(a\rightarrow \gamma\gamma) &= \frac{m_a^3}{4\pi} \,
|g^{\mathrm{eff}}_{a\gamma\gamma}|^2 \;,
\nonumber\\
\Gamma(a\rightarrow \gamma Z) &= \frac{m_a^3}{8\pi} \,
|g^{\mathrm{eff}}_{a\gamma Z}|^2 \left( 1 - \frac{m_Z^2}{m_a^2}
\right)^{\!\!3} ,
\nonumber\\
\Gamma(a\rightarrow ZZ) &= \frac{m_a^3}{4\pi} \,
|g^{\mathrm{eff}}_{aZZ}|^2 \left( 1 - \frac{4m_Z^2}{m_a^2}
\right)^{\!\!3/2} ,
\nonumber\\
\Gamma(a\rightarrow W^+W^-) &= \frac{m_a^3}{4\pi} \,
|g^{\mathrm{eff}}_{aWW}|^2 \left( 1 - \frac{4m_W^2}{m_a^2}
\right)^{\!\!3/2} ,
\end{align}
where couplings $g_{aVV}$ ($V = \gamma, Z, W$) are given by
eqs.~\eqref{ALP_boson_eff_couplings}.

We assume that the ALP Lagrangian has no terms like (gluon-phobic
ALPs)
\begin{equation}\label{ALP_gluon_lagrangian}
\mathcal{L}^a_\mathrm{g} = \frac{1}{f_a} \, c_{agg} G_{\mu\nu}
\tilde{G}^{\mu\nu} ,
\end{equation}
where $G_{\mu\nu}$ is the gluon strength tensor, and
$\tilde{G}_{\mu\nu}$ is the dual tensor. But as it takes place for
the vector bosons, the ALP coupling to top quarks
\eqref{ALP_top_lagrangian} would induce the ALP-gluon interaction
with the effective coupling \cite{Esser:2023}
\begin{equation}\label{ALP_gluon_eff_coupling}
g^{\mathrm{eff}}_{agg} = -\frac{\alpha_s}{8\pi f_a}
\,c_{{at\bar{t}}} \;.
\end{equation}
As a result, we get
\begin{equation}\label{ALP_decay_widths_into_gluons}
\Gamma(a\rightarrow gg) = \frac{m_a^3}{4\pi} \,
|g^{\mathrm{eff}}_{agg}|^2 .
\end{equation}
The total width of the ALP, $\Gamma_a$, is a sum of the partial
decay widths defined by eqs.~\eqref{ALP_decay_widths_into_top},
\eqref{ALP_decay_widths_into_bosons}, and
\eqref{ALP_decay_widths_into_gluons}.

If we take into account loop corrections from all fermions, as well
as $W$-boson loop, we get \cite{Bauer:2017} (see also
\cite{Esser:2023,Craig:2018})
\begin{equation}\label{ALP_photon_eff_coupling}
g^{\mathrm{eff}}_{a\gamma\gamma} = g_{a\gamma\gamma} - \sum_f
\frac{N_c^fQ_f^2 \alpha_{\mathrm{em}}}{4\pi}
\,\frac{c_{af\bar{f}}}{f_a} \,B_1 \!\left( \frac{4m_f^2}{p^2}
\right) + \frac{2\alpha_{\mathrm{em}}}{\pi} g_{aWW} B_2 \!\left(
\frac{4m_W^2}{p^2} \right) ,
\end{equation}
where $N_c^f$ denotes the color multiplicity of the fermion $f$, and
$Q_f$ is its electric charge (in units of $e$). The loop functions
are given by \cite{Bauer:2017}
\begin{align}\label{B1_B2}
B_1(x) & = 1 - xf^2\!(x) \;,
\nonumber \\
B_2(x) & = 1 - (x - 1)f^2\!(x) \;,
\end{align}
with
\begin{equation}\label{}
f(x) =
\left\{
  \begin{array}{ll}
    \displaystyle \arcsin\frac{1}{\sqrt{x}} ; & x \geq 1 \;, \\
 \displaystyle
   \frac{\pi}{2} + \frac{i}{2} \ln \frac{1 + \sqrt{1-x}}{1 - \sqrt{1-x}} ; &  x < 1 \;.
  \end{array}
\right.
\end{equation}
For $p^2 \gg m_f^2, m_W^2$ the loop functions have the properties
$B_1(4m_f^2/p^2) \simeq 1$, $B_2(4m_W^2/p^2) \simeq -(1/4)
\ln^2(p^2/m_W^2)$.

If we assume that $c_{at\bar{t}} \gg c_{af\bar{f}}$, where $f$
denotes ``light fermions'', than we get at $p^2 \gg m_V^2$
\begin{equation}\label{ALP_photon_eff_coupling_m}
g^{\mathrm{eff}}_{a\gamma\gamma} = g_{a\gamma\gamma} - \frac{
\alpha_{\mathrm{em}}}{3\pi} \,\frac{c_{at\bar{t}}}{f_a}  -
\frac{\alpha_{\mathrm{em}}}{2\pi} \,g_{aWW} \ln^2 \left(
\frac{p^2}{m_W^2} \right) ,
\end{equation}
where $g_{a\gamma\gamma}$ and $g_{aWW}$ are defined by
eqs.~\eqref{ALP_boson_couplings}.

The effective one-loop ALP-top coupling was obtained in
\cite{Bonilla:2021},
\begin{align}\label{ALP_top_eff_coupling_m}
\frac{c^{\mathrm{eff}}_{at\bar{t}}}{f_a} &=
\frac{c_{at\bar{t}}}{f_a} \left[ 1 +
\frac{\alpha_{\mathrm{em}}}{2\pi} D^{c_{att}} \right] +
\frac{\alpha_{\mathrm{em}}c_{at\bar{t}}}{2\pi f_a}
D^{c_{att}}_{\mathrm{mix}} \nonumber
\\
&+ \frac{2\alpha_{\mathrm{em}}}{\pi} \left[ g_{a\gamma\gamma}
D^{\gamma\gamma} + g_{a\gamma Z} D^{\gamma Z} + g_{aZZ} D^{ZZ} +
g_{aWW} D^{WW} \right] ,
\end{align}
where
\begin{equation}\label{D_att}
D^{c_{att}} = D^{c_{att}}_\gamma + D^{c_{att}}_Z + D^{c_{att}}_W +
D^{c_{att}}_h .
\end{equation}
Here $D^{c_{att}}_\gamma$, $D^{c_{att}}_Z$, $D^{c_{att}}_W$, and
$D^{c_{att}}_h$ are contributions from the one-loop exchange of the
photon, $Z$, $W$, and Higgs boson (vertex corrections), as well as
leg corrections. Their explicit expressions look like
\cite{Bonilla:2021}:
\begin{align}\label{D_att_gamma}
D^{c_{att}}_{\gamma} = &- \frac{4}{9} \bigg\{ 1 - \frac{\pi^2}{6} +
\ln\!\left( \frac{\lambda^2}{m_t^2} \right) \left[ 1 + i\pi +
\ln\!\left( \frac{m_t^2}{p^2} \right) \right]
\nonumber
\\
&+ \frac{1}{2} \left[ \ln\!\left( \frac{m_t^2}{p^2} \right) + i\pi
\right]^2 \bigg\} ,
\end{align}
where $\lambda$ is the infrared (IR) cutoff,
\begin{align}\label{D_att_Z}
D^{c_{att}}_Z &= \frac{1}{2c_w^2 s_w^2} \bigg\{
-\frac{m_t^2}{4m_Z^2} \left[ \ln\!\left( \frac{\Lambda^2}{p^2}
\right) + 2 + i\pi \right] \nonumber
\\
&- [(1/4) + (4/3)s_w^2 - (16/9)s_w^2] \ln\!\left(
\frac{m_t^2}{m_Z^2} \right)
\nonumber
\\
&- (1/3)s_w^2 [1 - (4/3)s_w^2] \bigg[ \frac{\pi^2}{3} + \ln\!\left(
\frac{m_t^2}{m_Z^2} \right) \left( \ln\!\left( \frac{m_t^2}{p^2}
\right) + i\pi \right)
\nonumber
\\
&- \left( \ln\!\left( \frac{m_t^2}{p^2} \right) + i\pi
\right)^{\!\!2} \bigg] \bigg\} \;,
\end{align}
where $\Lambda$ is the ultraviolet (UV) cutoff, and
\begin{equation}\label{D_att_W}
D^{c_{att}}_W = - \frac{m_t^2}{8s_w^2 m_W^2} \left[ \ln\!\left(
\frac{\Lambda^2}{m_t^2} \right) + 1 + i\pi \right] \;.
\end{equation}
\begin{equation}\label{D_att_h}
D^{c_{att}}_h = -\frac{m_t^2}{8\pi s_w^2 m_W^2} \left[ \ln\!\left(
\frac{\Lambda^2}{m_h^2} \right) +  \ln\!\left( \frac{p^2}{m_h^2}
\right) - 2 - i\pi \right] .
\end{equation}

Other corrections in \eqref{ALP_top_eff_coupling_m} are given by
\cite{Bonilla:2021}
\begin{equation}\label{D_att_mix}
D^{c_{att}}_{\mathrm{mix}} = -\frac{3m_t^2}{4s_w^2 m_W^2} \left[
\ln\!\left( \frac{\Lambda^2}{m_t^2} \right) + 2 + i\pi \right] ,
\end{equation}
and
\begin{equation}\label{D_gamma_gamma}
D^{\gamma\gamma} = \frac{2}{9} \left\{ 3\ln\!\left(
\frac{\Lambda^2}{m_t^2} \right) - 4 - \frac{2\pi^2}{3} - \frac{1}{2}
\left( \ln\!\left( \frac{m_t^2}{p^2} \right) + i\pi \right)^{\!\!2}
\right\} ,
\end{equation}
\begin{equation}\label{D_gamma_Z}
D^{\gamma Z} = \frac{1 - (8/3)s_w^2}{12c_w^2 s_w^2}  \left[
3\ln\!\left( \frac{\Lambda^2}{m_t^2} \right) - 4 \right] ,
\end{equation}
\begin{align}\label{D_ZZ}
D^{ZZ} &= \frac{1}{4c_w^2 s_w^2} \bigg\{ (1/4 - (2/3)s_w^2 +
(4/3)s_w^2) \left[ 3\ln\!\left( \frac{\Lambda^2}{m_t^2} \right) - 4
\right]
\nonumber
\\
&+ (4/9)s_w^2[1 - (4/3)s_w^2] \left[ \frac{2\pi^2}{3} + \frac{1}{2}
\left( \ln\!\left( \frac{m_t^2}{p^2} \right) + i\pi \right)^{\!\!2}
\right] \nonumber
\\
&+ \frac{1}{2} \left( \ln\!\left( \frac{m_t^2}{p^2} \right) + i\pi
\right) \bigg\} ,
\end{align}
\begin{equation}\label{D_WW}
D^{WW} = \frac{1}{8s_w^2} \left[ 3\ln\!\left(
\frac{\Lambda^2}{m_t^2} \right) - \ln\!\left( \frac{m_t^2}{p^2}
\right) - 3 + 3i\pi \right] .
\end{equation}

The third term in \eqref{D_att_gamma} exhibits a logarithmic
dependence on the IR cutoff $\lambda$. However, this divergence is
instead physically irrelevant and can be disregarded, as it will
exactly cancel with that stemming from soft and/or collinear photon
bremsstrahlung \cite{Bonilla:2021}. For instance, soft photon cross
sections are obtained by squaring the soft photon matrix elements,
summing over the photon polarizations, and integrating over the
photon energy $k_0 \leq \Delta E$ \cite{Denner:1993}. As in the
virtual corrections, the IR singularities are regularized by the
photon mass $\lambda$, and one gets a divergence of the type
$\ln\!(\Delta E^2/\lambda^2)$ \cite{Denner:1993}. Adding it to the
cross section for the corresponding basic process the IR
divergencies cancel. As a result, in the limit $\lambda \rightarrow
0$ the divergent term $\ln\!(\lambda^2/m_t^2)$ in
\eqref{D_att_gamma} will be replaced by the finite term
$\ln\!(\Delta E^2/m_t^2)$. The photon energy resolution at the CLIC
is expected to be $\Delta E/E = 20\%/\sqrt{E/\mathrm{GeV}} \oplus
1\%$ \cite{Ellis:2022,CLICdp:2018}. We find that $\Delta E = 22.75$
GeV (40.95 GeV) for $\sqrt{s} = 1.5$ TeV  (3 TeV).

As for the UV cutoff $\Lambda$ in Eqs.~\eqref{D_att_Z}-\eqref{D_WW},
following \cite{Bonilla:2021}, we assume that it should be of the
same order as $f_a$. In our numerical calculations we will put
$\Lambda = f_a$.

\section{Top pair production in $\gamma\gamma$ collision} %

A contribution of the ALP to the photon-induced top pair production
is defined by the process
\begin{equation}\label{top_pair_prod}
\gamma\gamma \rightarrow a \rightarrow t \bar{t} \;.
\end{equation}
The square of the amplitude of the process \eqref{top_pair_prod} is
equal to
\begin{equation}\label{ALP_amplitude}
|M_a(\hat{s})|^2 = \frac{4}{f_a^2}\frac{ [
g^{\mathrm{eff}}_{a\gamma\gamma} c^{\mathrm{eff}}_{at\bar{t}}
\,m_t]^2 \hat{s}^3 }{( \hat{s} - m_a^2)^2 + (m_a\Gamma_a)^2 } \;,
\end{equation}
where the effective couplings $g^{\mathrm{eff}}_{a\gamma\gamma}$ and
$c^{\mathrm{eff}}_{at\bar{t}}$ are defined by
eqs.~\eqref{ALP_photon_eff_coupling_m} and
\eqref{ALP_top_eff_coupling_m}, respectively, and $\hat{s}$ is the
collision energy squared of the backscattered photons. Note that the
coupling $g^{\mathrm{eff}}_{a\gamma\gamma}$ is proportional to
$f_a^{-1}$, see eqs.~\eqref{ALP_boson_eff_couplings},
\eqref{ALP_boson_couplings}.

Let $E_e$ be the energy of the electron beam, $4E_e^2 = s$, and
$E_{\gamma, 1}$, $E_{\gamma, 2}$ are the energies of the colliding
CB photons. Then the differential cross section of the process
\eqref{top_pair_prod} depends on spectra of the CB photons
$f_{\gamma/e}(x_i)$ ($i = 1,2$) and total amplitude $M = M_a +
M_{\mathrm{SM}}$ \cite{I_K:2021_1},
\begin{equation}\label{diff_cs_in}
\frac{d\sigma}{d\cos\theta} = \frac{\beta}{128\pi\hat{s}}
\int\limits_{x_{\min}}^{x_{\max}} \!dx_1 \,f_{\gamma/e}(x_1)
\int\limits_{x_{\min}}^{x_{\max}} \! dx_2 \,f_{\gamma/e}(x_2)
\,|M(\hat{s}, \cos\theta)|^2 ,
\end{equation}
where $\hat{s} = 4E_{\gamma, 1}E_{\gamma, 2} = sx_1 x_2$, $x_1 =
E_{\gamma, 1}/E_e$ and $x_2 = E_{\gamma, 2}/E_e$ are the energy
fractions of the CB photon beams, $x_{\max} = 0.83$, and $\beta = 1
- 4m_t^2/\hat{s}$. After changing variables, $x = x_2$, $\tau = x_1
x_2$ the differential cross section \eqref{diff_cs_in} takes the
form
\begin{equation}\label{diff_cs}
\frac{d\sigma}{d\cos\theta} = \frac{\beta}{128\pi\hat{s}}
\int\limits_{\tau_{\min}}^{x_{\max}^2} \!\!d\tau
\!\!\int\limits_{\tau/x_{\max}}^{x_{\max}} \!\!\frac{dx}{x}
\,f_{\gamma/e}(x) \,f_{\gamma/e}\!\left( \frac{\tau}{x} \right)
|M(\hat{s}, \cos\theta)|^2 ,
\end{equation}
where $\tau_{\min} = 4m_t^2/s$.%
\footnote{As was mentioned in Introduction, we consider the CLIC
energies satisfying the condition $0.83 \sqrt{s} > 2m_t$.}

\begin{figure}[htb]
\begin{center}
\includegraphics[scale=0.5]{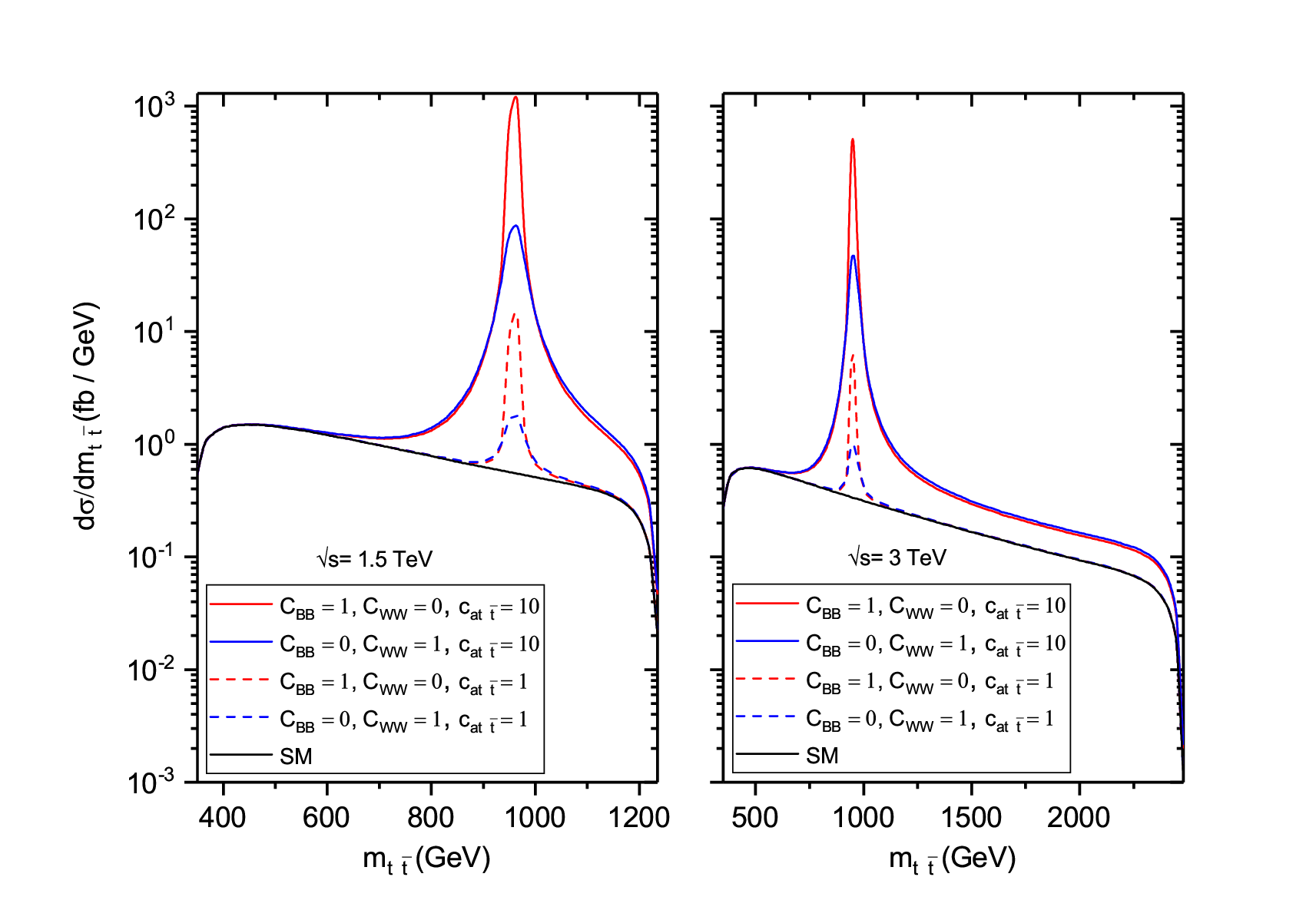}
\caption{The differential cross sections of the collision
$\gamma\gamma \rightarrow t\bar{t}$ at the CLIC versus invariant
mass of the top pair. The parameters of the ALP are: $m_a = f_a = 1$
TeV.}
\label{fig:MDCSBf1_C}
\end{center}
\end{figure}

\begin{figure}[htb]
\begin{center}
\includegraphics[scale=0.5]{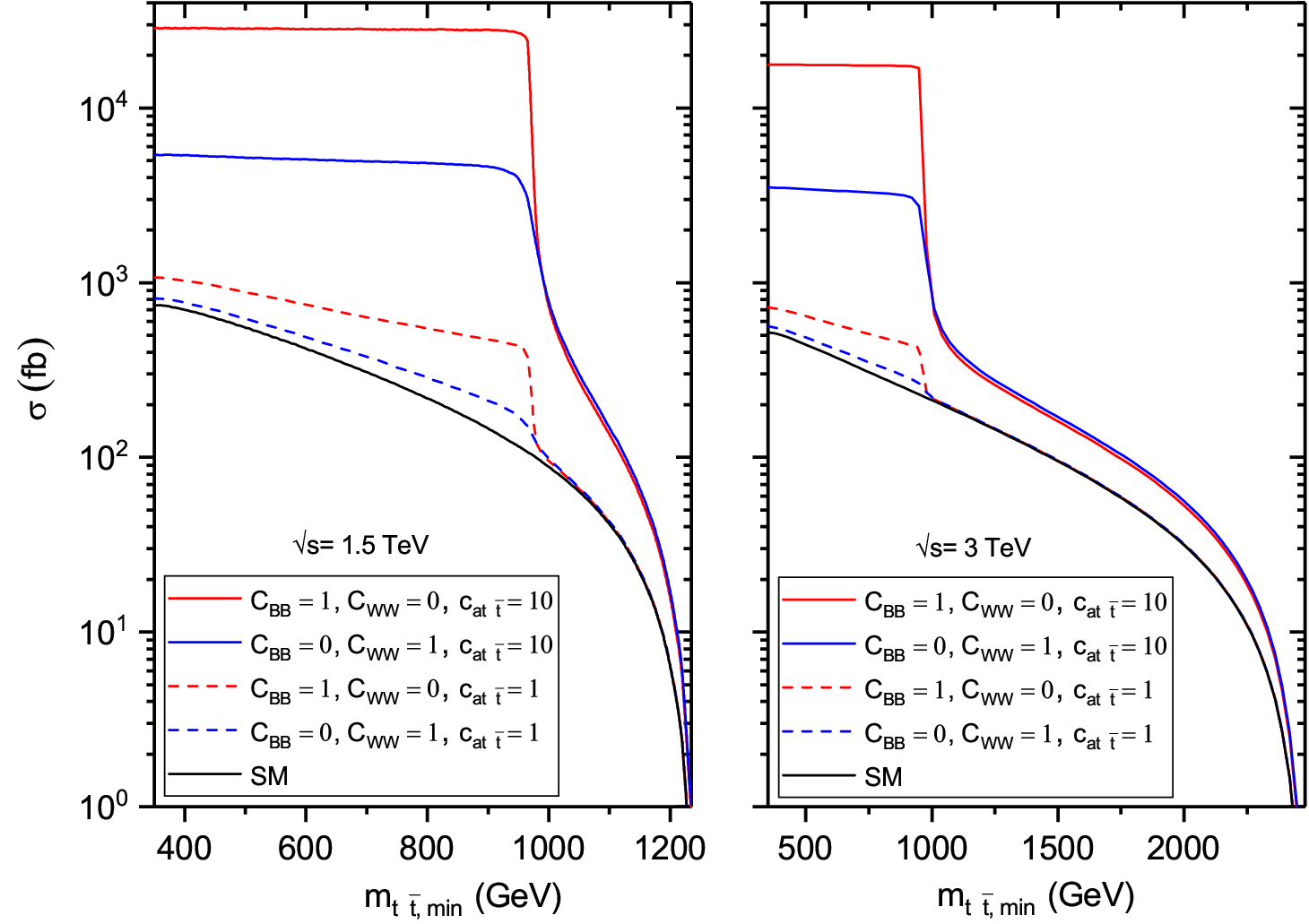}
\caption{The total cross sections of the collision $\gamma\gamma
\rightarrow t\bar{t}$ at the CLIC versus minimal value of the
invariant mass of the top pair. The parameters of the ALP are: $m_a
= f_a = 1$ TeV.}
\label{fig:MCUTBR1f1_C}
\end{center}
\end{figure}

The differential cross sections of this process for two values of
the CLIC energy $s$ is shown in Fig.~\ref{fig:MDCSBf1_C} as
functions of the invariant mass of the top-quark pair
$m_{t\bar{t}}$.
In Fig.~\ref{fig:MCUTBR1f1_C} the dependence of the total cross
sections on the minimal value of $m_{t\bar{t}}$ are presented. We
applied the cuts on the rapidity and transverse momentum of the
outgoing top-quarks, $|\eta| < 2.5$ and $p_T > 30$ GeV,
correspondingly.

Using these cross sections, we can obtain exclusion bounds on the
scale $f_a^{-1}$ depending on the ALP mass $m_a$ and coefficient
$c_{at\bar{t}}$ in the $\gamma\gamma \rightarrow t \bar{t}$
scattering at the CLIC operating in the $\gamma\gamma$ mode. To
derive these bounds, we apply the formula for the statistical
significance $SS$ \cite{SS}
\begin{equation}\label{SS_def}
SS = \sqrt{2[(S - B\,\ln(1 + S/B)]} \;.
\end{equation}
Here $S$ denotes the number of signal events, while $B$ means the
number of background (SM) events. The number of signal events is
calculated via $\sigma_{\mathrm{ALP}}(\gamma \gamma \rightarrow t
\bar{t}) \mathrm{B}(t \rightarrow W^+ b) \mathrm{B}(\bar{t}
\rightarrow W^- \bar{b})\mathrm{B}(W^+ \rightarrow
\mathrm{hadrons})$ $\mathrm{B}(W^- \rightarrow \mathrm{hadrons})$,
where $\sigma_{\mathrm{ALP}}(\gamma \gamma \rightarrow t \bar{t})$
is the total cross section of the $\gamma \gamma \rightarrow a
\rightarrow t \bar{t}$ process whose amplitude squared is given by
\eqref{ALP_amplitude}. So, we assume that both $W$-bosons decay
hadronically. The values of the top mass $m_t$ and branching ratios
are taken from \cite{PDG:2024}.

As for a SM background, we consider the LO process $\sigma(\gamma
\gamma \rightarrow W^+b W^-\bar{b})$ with the cuts $|m_{W^+ b} -
m_t| < 5$ GeV and $|m_{W^-\bar{b}} - m_t| < 5$ GeV applied,  using
CalcHep package \cite{CalcHep}, where $m_{W^+ b}(m_{W^- \bar{b}})$
is an invariant mass of the $W^+ b(W^- \bar{b})$ system.

The higher QCD corrections to $t\bar{t}$ cross section, as shown in
\cite{Chen:2024} (see also \cite{Ma:2024}), are more significant for
the process with smaller $e^+e^-$ collision energies $\sqrt{s}$. But
they decrease monotonically as $\sqrt{s}$ grows, and already at
$\sqrt{s} = 1000$ GeV, the NLO and NNLO corrections reduce to only
5\% and 0.3\%, respectively \cite{Chen:2024}. The NNNLO QCD
corrections are estimated to be of approximately 0.1\% only
\cite{Chen:2024_2}. The QED correction to the $e^+e^- \rightarrow
t\bar{t}$ cross section \cite{Fleischer:2003,Khiem:2013} is also
dominant in the low energy region. In the high energy region it is
much smaller ($\sim 5\%$ at 1 TeV ), see, for instance, Fig.~2 in
\cite{Khiem:2013}.

\begin{figure}[htb]
\begin{center}
\hspace*{-0.4cm}
\includegraphics[scale=0.5]{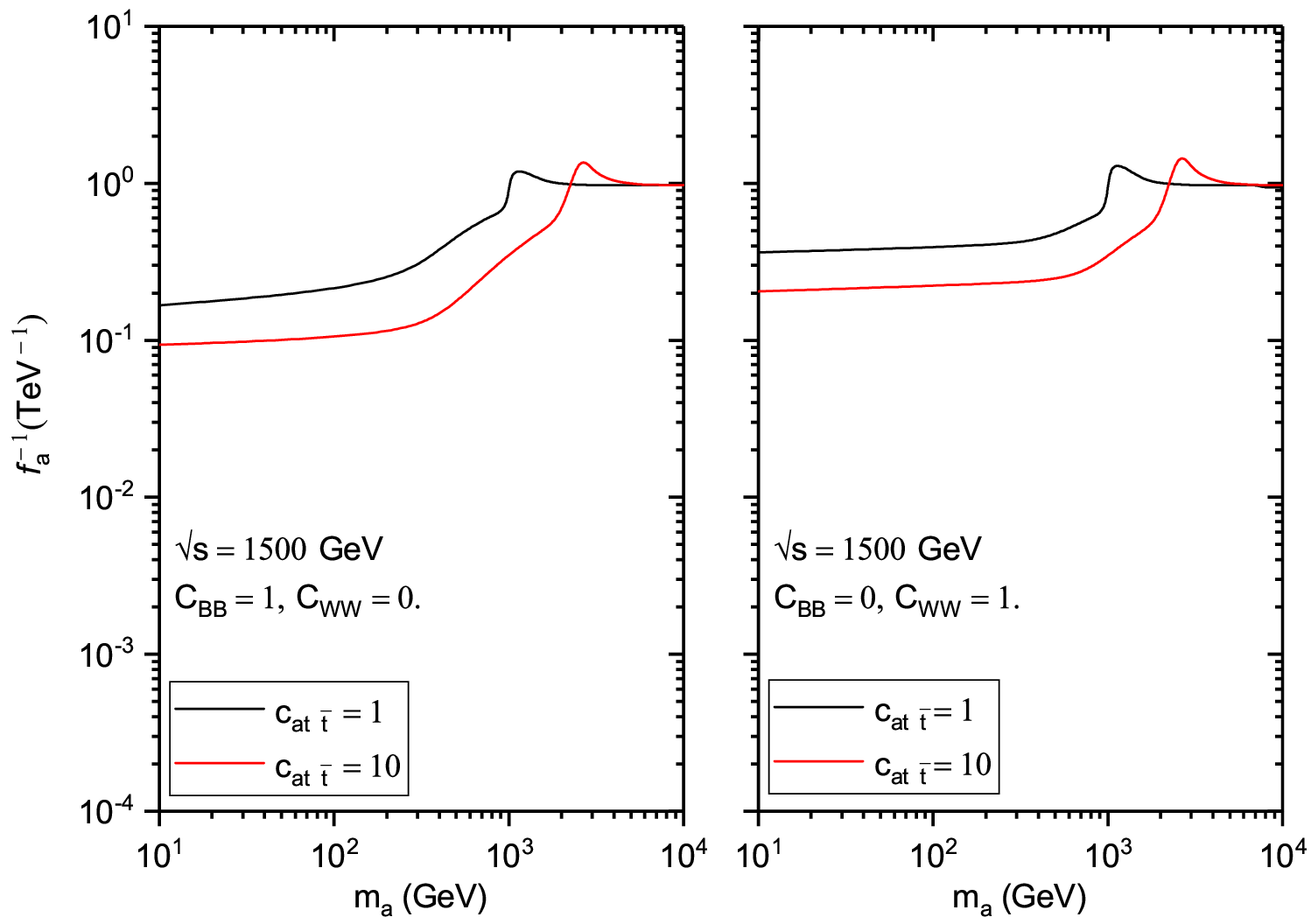}
\caption{The exclusion bounds on the scale $f_a$ via ALP mass $m_a$.
The collision energy of the CLIC is equal to 1.5 TeV.}
\label{fig:SSE750ALPTOP}
\end{center}
\end{figure}

We define the regions $SS \leqslant 1.645$ as the regions that can
be excluded at the 95\% C.L. To reduce the SM background, we
additionally to our previous cuts impose the cut $m_{t\bar{t}} >
800$ GeV. Our results for $\sqrt{s} = 1.5$ TeV are shown in
Fig.~\ref{fig:SSE750ALPTOP}. We used the the integrated luminosities
$L = 2.5$ ab$^{-1}$ for $\sqrt{s} = 1.5$ TeV and $L = 5$ ab$^{-1}$
for $\sqrt{s} = 3$ TeV.

To understand various features of the curves in Fig.~3, let us
analyze the ALP amplitude squared $|M_a|^2$ (30). In the region $10
\mathrm{\ GeV} \leqslant m_a \lesssim 300$ GeV we have $(s - m_a^2)
\simeq s \gg (m_a \Gamma_a)^2$, and we get
\begin{equation}\label{ALP_ampl_appr_1}
|M_a(\hat{s})|^2 \sim \frac{ m_t^2 \hat{s}}{f_a^4} \;.
\end{equation}
It means that the ALP cross is almost independent of $m_a$.
Correspondingly, in the above mentioned region both curves in Fig.~3
are slowly changing with $m_a$.

In the intermediate region $300 \mathrm{\ GeV} < m_a \lesssim 1$ TeV
the amplitude squared becomes smaller as $m_a$ grows due to the
factor $(m_a\Gamma_a)^2 \approx (m_a^4/(\pi f_a^2))^2$ in the
denominator of eq.~(30). Correspondingly, the bound on the scale
$1/f_a$ becomes weaker with an increase of $m_a$.

In the region around $m_a \simeq 1$ TeV we see a small asymmetric
resonance.

Finally, in the region $2 \mathrm{\ TeV} \lesssim m_a \leqslant 10$
TeV, we have the inequality $m_a\Gamma_a \gg m_a^2 - s$, that
results in an approximate formula
\begin{equation}\label{ALP_ampl_appr_2}
|M_a(\hat{s})|^2 \sim \frac{m_t^2 \hat{s}^3}{m_a^8} \;.
\end{equation}
It means that the number of signal events is almost independent on
$f_a$. That is why, in the region of large $m_a$ the bound on
$1/f_a$ is constant and equal to its value in the resonance region.
This is in agreement with the features of the curves in Fig.~3.

In the next Fig.~\ref{fig:SSE1500ALPTOP} the exclusion bounds for
the CLIC energy of $\sqrt{s} = 3$ TeV are given.
\begin{figure}[htb]
\begin{center}
\hspace*{-0.4cm}
\includegraphics[scale=0.5]{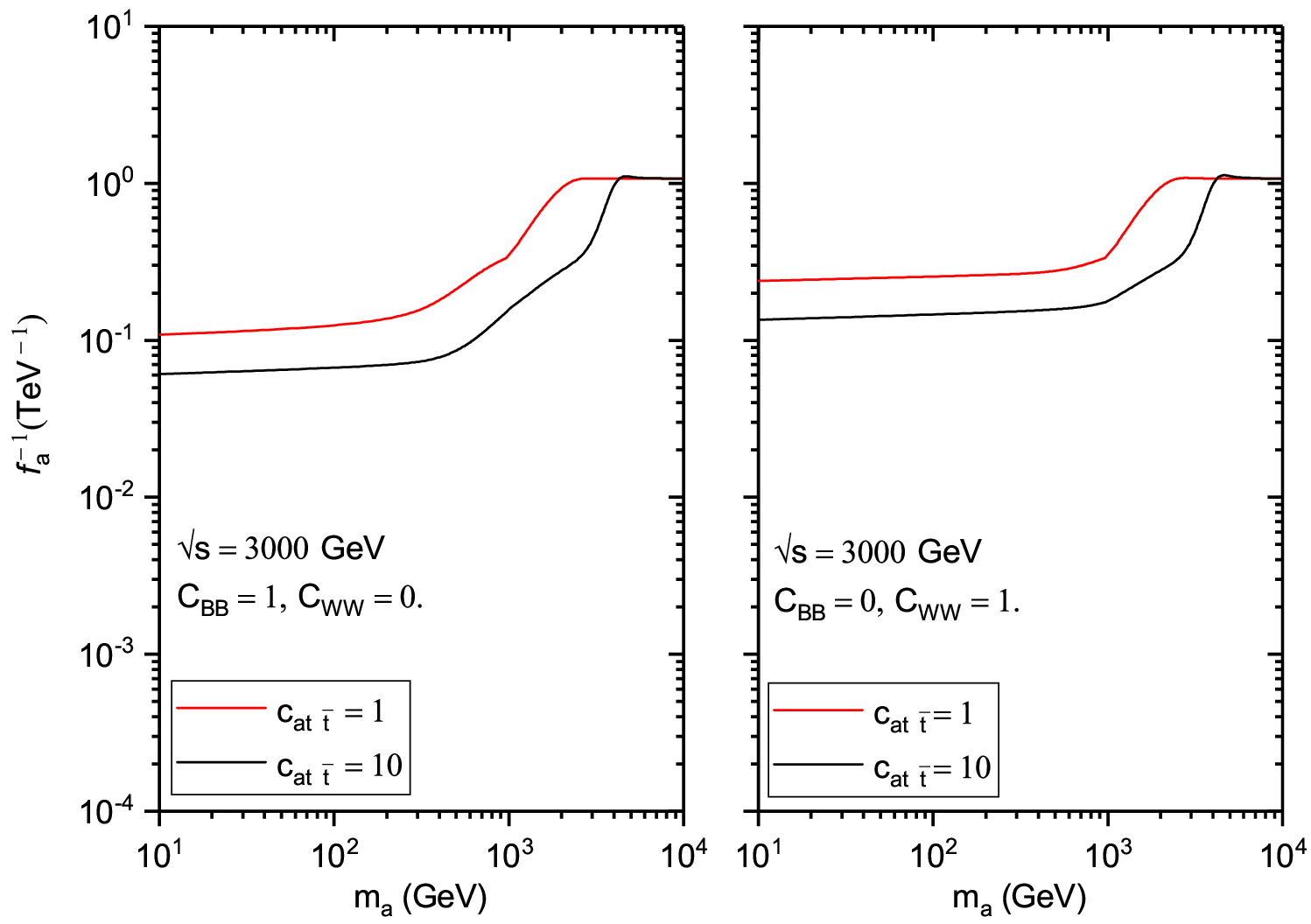}
\caption{The same as in Fig.~\ref{fig:SSE750ALPTOP}, but for the
CLIC energy of 3 TeV.}
\label{fig:SSE1500ALPTOP}
\end{center}
\end{figure}
The limits for the ALP-top quark coupling $c_{at\bar{t}}/f_a$ are
collected in Tabs.~1 and 2. As one can see from these tables, the
bounds on $c_{at\bar{t}}/f_a$ become weaker, as the ALP mass $m_a$
grows. For $m_a = 10$ GeV they are equal to $1.66 \times
10^{-1}$TeV$^{-1}$ and $1.09 \times 10^{-1}$ TeV$^{-1}$, for the
CLIC energy of 1.5 TeV and 3 TeV, respectively. These values of
$c_{at\bar{t}}/f_a$ are realized for the gauge couplings set $C_{BB}
= 1, C_{WW} = 0$. If one takes $C_{BB} = 0, C_{WW} = 1$, the bounds
become approximately twice weaker for small $m_a$. But for $m_a \geq
1$ TeV they are almost independent of the ALP-gauge boson couplings
$C_{BB}$ and $C_{WW}$.
\begin{table}
\begin{center}
\begin{tabular}{|c||c|c|}
  \hline
  \multicolumn{1}{|c||}{} & \multicolumn{2}{|c|}{$c_{at\bar{t}}/f_a$, TeV$^{-1}$} \\
  \hline
  $m_a$, GeV & $C_{BB} = 1, C_{WW} =
  0$ & $C_{BB} = 0, C_{WW} = 1$ \\
  \hline\hline
  10  & $1.66 \times 10^{-1}$ & $3.61 \times 10^{-1}$ \\
  \hline
  100 & $2.11 \times 10^{-1}$ & $3.89 \times 10^{-1}$  \\
  \hline
  1000 & $6.69 \times 10^{-1}$ & $6.72 \times 10^{-1}$ \\
  \hline
  10000 & $9.71 \times 10^{-1}$ & $9.72 \times 10^{-1}$ \\
  \hline
\end{tabular}
\end{center}
\caption{The bounds on the ALP-top quarks couplings vs. ALP mass
$m_a$ which can be probed at the CLIC with the collision energy of
1.5 TeV.}
\label{tab:1.5TeV}
\end{table}
\begin{table}
\begin{center}
\begin{tabular}{|c||c|c|}
  \hline
  \multicolumn{1}{|c||}{} & \multicolumn{2}{|c|}{$c_{at\bar{t}}/f_a$, TeV$^{-1}$} \\
  \hline
  $m_a$, GeV & $C_{BB} = 1, C_{WW} =
  0$ & $C_{BB} = 0, C_{WW} = 1$ \\
  \hline\hline
  10  & $1.09 \times 10^{-1}$ & $2.38 \times 10^{-1}$  \\
  \hline
  100 & $1.18\times 10^{-1}$ & $2.46 \times 10^{-1}$ \\
  \hline
  1000 & $3.38\times 10^{-1}$ & $3.39 \times 10^{-1}$  \\
  \hline
  10000 & \!\!\!$10.71 \times 10^{-1}$ & \!\!\!$10.73 \times 10^{-1}$ \\
  \hline
\end{tabular}
\end{center}
\caption{The same as in Tab.~1, but for the CLIC energy of 3 TeV.}
\label{tab:3TeV}
\end{table}

\section{Comparison with other searches of the ALP-top coupling} %

In \cite{Esser:2023} the coupling of a light ALP to the top quarks
was examined. Both the direct probe to this coupling in the
associated production of a top-pair with an ALP and the indirect
probe through loop-induced gluon fusion to an ALP leading to top
pairs were studied. The constraint at 96\% C.L. following from the
ATLAS search looks like (here and in what follows $c_t \equiv
c_{at\bar{t}}$)
\begin{equation}\label{ALP-top_Esser_1}
\left| \frac{c_t}{f_a} \right| < 1.81 \mathrm{\ TeV}^{-1} .
\end{equation}
The ALP-mediated di-boson production via top coupling results in the
bound
\begin{equation}\label{ALP-top_Esser_2}
\left| \frac{c_t}{f_a} \right| < 44.4 \mathrm{\ TeV}^{-1} .
\end{equation}
The ATLAS measurement of top quark pair production with a high-$p_T$
top quark was also used in \cite{Esser:2023} to get the limit
\begin{equation}\label{ALP-top_Esser_3}
\left| \frac{c_t}{f_a} \right| < 5.9 \mathrm{\ TeV}^{-1} .
\end{equation}
From lower-energy precision NA62 and BaBar measurements of rare Kaon
and B-meson decays the following bounds are obtained
\cite{Esser:2023}
\begin{align}\label{ALP-top_Esser_4}
\left| \frac{c_t}{f_a} \right| &< 2.8 \times 10^{-1} \mathrm{\
TeV}^{-1} , \
(\mathrm{K\ decays}, m_a < 110 \mathrm{\ MeV}. \ m_a \in [160, 260] \mathrm{\ MeV}) \;,
\nonumber
\\
\left| \frac{c_t}{f_a} \right| &< 1.1 5 \times 10^{-3} \mathrm{\
TeV}^{-1} , \quad (\mathrm{B\ decays},  m_a \lesssim 5 \mathrm{\
GeV}) \;.
\end{align}

In \cite{Phan:2024} an in-depth analysis of axions and axion-like
particles in top-pair production at the LHC was presented. Bounds on
the ALP-top coupling from the top-antitop total cross section and
from the differential distributions are obtained,
\begin{equation}\label{ALP-top_Phan_1}
\left| \frac{c_t}{f_a} \right| < 7.0 \mathrm{\ TeV}^{-1} \;, \quad
(0 < m_a < 200 \mathrm{\ GeV}) \;.
\end{equation}
From the data on the top-antitop production with di-muon resonances
from the ATLAS \cite{ATLAS:ALP-top} and CMS \cite{CMS:ALP-top}
collaborations, the following limits are obtained
\begin{align}\label{ALP-top_Phan_2}
\left| \frac{c_t}{f_a} \right| &< 1 \mathrm{\ TeV}^{-1} , \qquad
(\mathrm{LHC \ 150\ fb}^{-1}) \;,
\nonumber
\\
\left| \frac{c_t}{f_a} \right| &< 0.1 \mathrm{\ TeV}^{-1} , \quad
(\mathrm{HL-LHC \ 3\ ab}^{-1}) \;.
\end{align}
From the study of a pair top production at the LHC it is found in
\cite{Bruggisser:2024} that
\begin{equation}\label{ALP-top_Bruggisser}
\left| \frac{c_t}{f_a} \right| < 4 \mathrm{\ TeV}^{-1} , \quad (m_a
= 300 \mathrm{\ GeV}) \;.
\end{equation}
A similar bound,
\begin{equation}\label{ALP-top_Blasi}
\left| \frac{c_t}{f_a} \right| < 6 \mathrm{\ TeV}^{-1} , \quad (10
\mathrm{\ GeV } < m_a < 200 \mathrm{\ GeV}) \;,
\end{equation}
is obtained in \cite{Blasi:2024}.
In \cite{Rygaad:2023} it is found
that the (HL-)LHC can probe effective top-quark couplings as small
as
\begin{equation}\label{ALP-top_Rygaad}
\frac{c_t}{f_a} = 0.03 \ (0.002) \mathrm{\ TeV}^{-1} , \quad (2m_\mu
< m_a < 2 \mathrm{\ GeV}) \;,
\end{equation}
assuming a cross section of 1 fb, with data corresponding to an
integrated luminosity of 150 fb$^{-1}$ (3 ab$^{-1}$).

Finally, the partial-wave unitarity bound on the  ALP-top coupling
looks like \cite{Brivio:2021}
\begin{equation}\label{unitariry_bounds}
\left| \frac{c_t}{f_a} \right| < 30 \left(
\frac{\mathrm{TeV}}{\sqrt{s}} \right) \mathrm{TeV}^{-1} \;.
\end{equation}
Then for the 3 TeV CLIC we get a the constraint $|c_t/f_a| < 10
\mathrm{\ TeV}^{-1}$. Thus, the unitarity is not violated in the
region of the ALP-top couplings studied in the present paper.

\begin{figure}[htb]
\begin{center}
\hspace*{-0.4cm}
\includegraphics[scale=0.85]{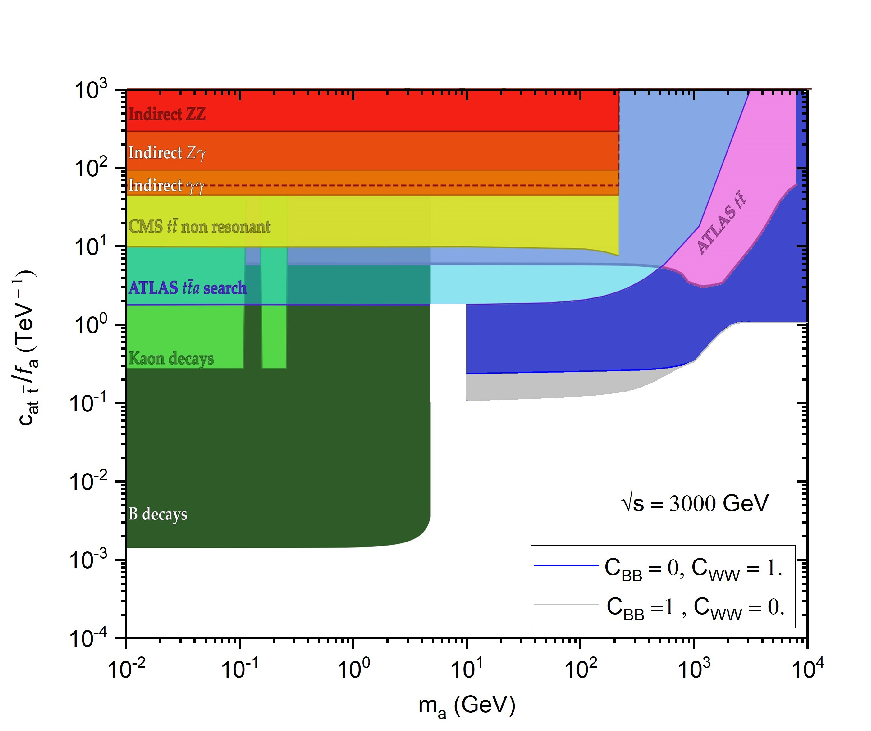}
\caption{The summary figure with the existing bounds on the ALP-top
coupling \protect{\cite{Esser:2023}} in comparison with our
predictions for the 3 TeV CLIC given by blue and gray regions.}
\label{fig:SSE1500_all}
\end{center}
\end{figure}

A summary plot is presented in Fig.~\ref{fig:SSE1500_all}
As one can see, in the region $10 \mathrm{\ GeV} < m_a < 1 $ TeV our
bounds on the ALP-top coupling $c_{at\bar{t}}/f_a$ are an order of
magnitude stronger than the constraints obtained by a direct ATLAS
$t\bar{t}a$ search and ATLAS measurement of top quark pair
production with a high-$p_T$ top quarks.

\section{Conclusions and discussion} %

In the present paper, we have studied the contribution of the ALPs
to the top pair production at the CLIC via the scattering of the
Compton backscattered photons. The collision energies of 1.5 TeV and
3 TeV are considered. We have presented both differential and total
cross sections depending on the invariant mass of the top quarks
pair. The bounds on the ALP-top quark coupling $c_{at\bar{t}}/f_a$
are obtained as functions of the ALP mass $m_a$ in the mass range 10
GeV -- 10 TeV. We have found that these constraints become less
stringent with an increase of $m_a$. The best  bounds on
$c_{at\bar{t}}/f_a$ for $m_a = 10$ GeV are equal to $0.17$
TeV$^{-1}$ and $0.11$ TeV$^{-1}$ for the CLIC energy of 1.5 TeV and
3 TeV, respectively. These bounds are compatible with constraints on
$c_{at\bar{t}}/f_a$ obtained for the HL-LHC with 3 ab$^{-1}$. We can
conclude that the CLIC operating in the $\gamma\gamma$ mode has a
great physical potential in searching for the ALP-top quarks
coupling.

The scale $f_a = 1\div 10$ TeV is comparable with the CLIC
center-off-mass energy. It means that above a scale of few TeV the
EFT needs more general QFT (UV completion). An UV complete theory
must be well-defined at arbitrarily high energies. Such a completion
could be, for instance, a strong dynamics with the ALP (and,
possibly, the top quark) being a composite state.

There is a class of composite Higgs models predicting the existence
of the ALP \cite{Ferretti:2014}-\cite{Ferretti:2025}. A composite
Higgs boson is accompanied by pseudoscalar light states generated by
the same dynamics. These are models based on underlying
four-dimensional gauge theories with hyperfermions, where the top
arises as fermionic composite state as in partial compositeness
model \cite{Kaplan:1991}. Recently a novel class of dark matter
candidates in the form of a heavy composite ALP was proposed
\cite{Carenza:2024}. It was argued that such a composite ALP emerges
as a bound state, the dark glueball, due to confinement in a pure
Yang-Mills dark sector.

The string theory (for instance, heterotic string or type IIB string
theory with D-branes) also suggests the simultaneous presence of
many pseudoscalar fields whose masses are logarithmically
hierarchical. Conversely, the presence of such a plenitude of axions
or ALPs (an ``axiverse'' \cite{Arvanitaki:2010}) would be evidence
for string theory. They arise due to the topological complexity of
the extra-dimensional manifold and is ad hoc in a theory with just
the four dimensions. Pseudoscalar fields with axion-like properties
generically arise in string theory compactifications as Kaluza-Klein
zero modes of antisymmetric tensor fields
\cite{Green:1987,Svrcek:2006}. The ALP-photon coupling was modelled
in a range of string axiverse scenarios
\cite{Halverson:2019,Gendler:2023}.

In Randall-Sundrum-type models \cite{Haque:2024}, a 4D effective
axion (ALP) can be obtained as the phase of a complex bulk scalar or
5-th component of a $U(1)$ bulk gauge field with appropriate
boundary conditions.

To summarize, the main goal of this article is to get bounds on the
ALP-top coupling. In addition, the process we examine allows one to
see an effect of the ALP-gauge boson coupling. When all these are
considered, we think that the gamma-gamma mode of the CLIC is one of
the most suitable process to achieve our goal.

\section*{Acknowledgements} %

This work is partially supported by the Scientific Research Project
Fund of Sivas Cumhuriyet University under project number
``F-2024-718''.




\end{document}